\documentclass[prb,twocolumn,superscriptaddress,preprintnumbers,a4paper,amsmath,amssymb,showpacs,floatfix]{revtex4}
\usepackage{graphicx}
\usepackage{bm}

\newcommand{\cfa}{CaFe$_2$As$_2$}
\newcommand{\afa}{$AE$Fe$_2$As$_2$}

\begin{document}


\title{Evidence from neutron diffraction for superconductivity in the
    stabilized tetragonal phase of \cfa under uniaxial pressure}

\author{K.~Proke\v{s}}
\email{prokes@helmholtz-berlin.de}\affiliation{Helmholtz-Zentrum
Berlin f\"{u}r Materialien und Energy, Glienicker Str.~100, 14109
Berlin, Germany}

\author{A.~Kreyssig}
\affiliation{
Ames Laboratory, U.S. DOE, Iowa State University, Ames, Iowa 50011, USA
 }%
\affiliation{Department of Physics and Astronomy, Iowa State
University, Ames, Iowa 50011, USA
 }%

\author{B.~Ouladdiaf}
\affiliation{
Institut Laue-Langevin, 38042 Grenoble Cedex, France
}%

\author{D.~K.~Pratt}
\affiliation{
Ames Laboratory, U.S. DOE, Iowa State University, Ames, Iowa 50011, USA
 }%
\affiliation{Department of Physics and Astronomy, Iowa State
University, Ames, Iowa 50011, USA
 }%

\author{N.~Ni}
\affiliation{
Ames Laboratory, U.S. DOE, Iowa State University, Ames, Iowa 50011, USA
 }%
\affiliation{Department of Physics and Astronomy, Iowa State
University, Ames, Iowa 50011, USA
 }%

\author{S.~L.~Bud'ko}
\affiliation{
Ames Laboratory, U.S. DOE, Iowa State University, Ames, Iowa 50011, USA
 }%
\affiliation{Department of Physics and Astronomy, Iowa State
University, Ames, Iowa 50011, USA
 }%

\author{P.~C.~Canfield}
\affiliation{
Ames Laboratory, U.S. DOE, Iowa State University, Ames, Iowa 50011, USA
 }%
\affiliation{Department of Physics and Astronomy, Iowa State
University, Ames, Iowa 50011, USA
 }%

\author{R.~J.~McQueeney}
\affiliation{
Ames Laboratory, U.S. DOE, Iowa State University, Ames, Iowa 50011, USA
 }%
\affiliation{Department of Physics and Astronomy, Iowa State
University, Ames, Iowa 50011, USA
 }%

\author{D.~N.~Argyriou}
\affiliation{Helmholtz-Zentrum Berlin f\"{u}r Materialien und
Energy, Glienicker Str.~100, 14109 Berlin, Germany}

\author{A.~I.~Goldman}
\affiliation{
Ames Laboratory, U.S. DOE, Iowa State University, Ames, Iowa 50011, USA
 }%
\affiliation{Department of Physics and Astronomy, Iowa State
University, Ames, Iowa 50011, USA
 }%

\date{\today}
\begin{abstract}
\cfa~single crystals under  uniaxial pressure applied along  the $c$
axis exhibit the coexistence of several structural  phases at low
temperatures.  We show that the room temperature tetragonal phase is
stabilized at low temperatures for pressures above 0.06~GPa, and its
weight fraction attains a maximum in the region where
superconductivity is observed under applied uniaxial pressure.
Simultaneous resistivity measurements strongly suggest that this
phase is responsible for the superconductivity in \cfa~found below
10~K in samples subjected to non-hydrostatic pressure conditions.

\end{abstract}

\maketitle Since their discovery, both the 1111 oxypnictide
\cite{Kamihara:2006} and 112 iron arsenide \cite{Rotter:2008}
superconducting families have undergone intensive scrutiny,
particularly with respect to relationships between structure,
magnetism, composition and superconductivity ($SC$)
\cite{Ishida:2009,Canfield:2009}. The parent $R$FeAsO ($R$ = rare
earth) and \afa~($AE$ = Ca, Sr, Ba) compounds are not
superconductors at ambient pressure, but undergo structural and
antiferromagnetic ($AF$) transitions that are, at least in some
instances, strongly coupled
\cite{Ishida:2009,Canfield:2009,Krellner:2008,Goldman:2008}. Upon
chemical doping \cite{Rotter:2008,Ishida:2009} or under pressure
\cite{Takahashi:2008,Torikachvili:2008a}, the structural and
magnetic transitions are suppressed and $SC$ is observed with
$T_{C}$ as high as 55~K \cite{Ren:2008}.

One of the most interesting anomalies in the \afa~family is found in
\cfa~under pressure as discussed in a recent review
\cite{Canfield:2009}. At ambient pressure, \cfa~undergoes a first
order transition from a high temperature tetragonal ($T$) phase
(ThCr$_2$Si$_2$ structure) to a structure with orthorhombic ($O$)
symmetry at $T_{TO}$ = 172 K \cite{Ni} concomitant with an $AF$ transition \cite{Goldman:2008}. Upon the application of
modest pressures, using liquid media self-clamping cells, the
structural and $AF$ transitions were rapidly suppressed and $SC$ was
observed for $P$ $\geq$ 0.23~GPa and $T$ $\leq$ 12 K
\cite{Torikachvili:2008a,Park:2008}. $SC$ has also been observed in
electrical resistance measurements of samples under uniaxial
pressure \cite{Torikachvili:2009a}.

Neutron powder diffraction measurements, using a He gas pressure
cell to ensure hydrostatic pressure conditions, revealed a volume-collapsed tetragonal ($cT$) phase in this pressure range, below
$\thickapprox$ 100~K \cite{Kreyssig:2008a}. Although the onset of $SC$ seemed to be
closely related to the appearance of the $cT$ phase, more
recent transport measurements under hydrostatic pressure conditions
(He-gas cell) have revealed that neither the ambient pressure $O$
phase (below $T_{TO}$) nor the $cT$ phase support $SC$
\cite{Yu:2009}. These measurements along with an extended structural
study by single crystal neutron diffraction \cite{Goldman:2009},
demonstrated that the electronic, magnetic and structural
transitions are sharp and clearly defined under hydrostatic
pressure. Measurements done using a frozen liquid medium, in
contrast, manifest a significant non-hydrostatic component upon the
transition to the $cT$ phase resulting in a low temperature
multi-crystallographic-phase state that includes both the $O$ and
$cT$ phases among, perhaps, other as yet unidentified phases. This
is consistent with reports of the coexistence between static
magnetic order and $SC$ as inferred from $\mu$SR experiments
\cite{Goko} and recent NMR experiments \cite{Baek}. Nevertheless,
the puzzle remains: Which phase(s) is(are) responsible for $SC$ in
\cfa~under pressure? Does the orthorhombic phase support both
superconductivity and magnetic ordering or, as speculated in Ref.
~\cite{Canfield:2009}, is $SC$ associated with some residual
untransformed $T$ phase? Is $SC$ to be found in this, as yet
undiscovered phase at the boundary between the $O$ and $cT$ phases
\cite{Yu:2009,Torikachvili:2009a,Lee:2009}?

\begin{figure}
\includegraphics*[scale=0.34]{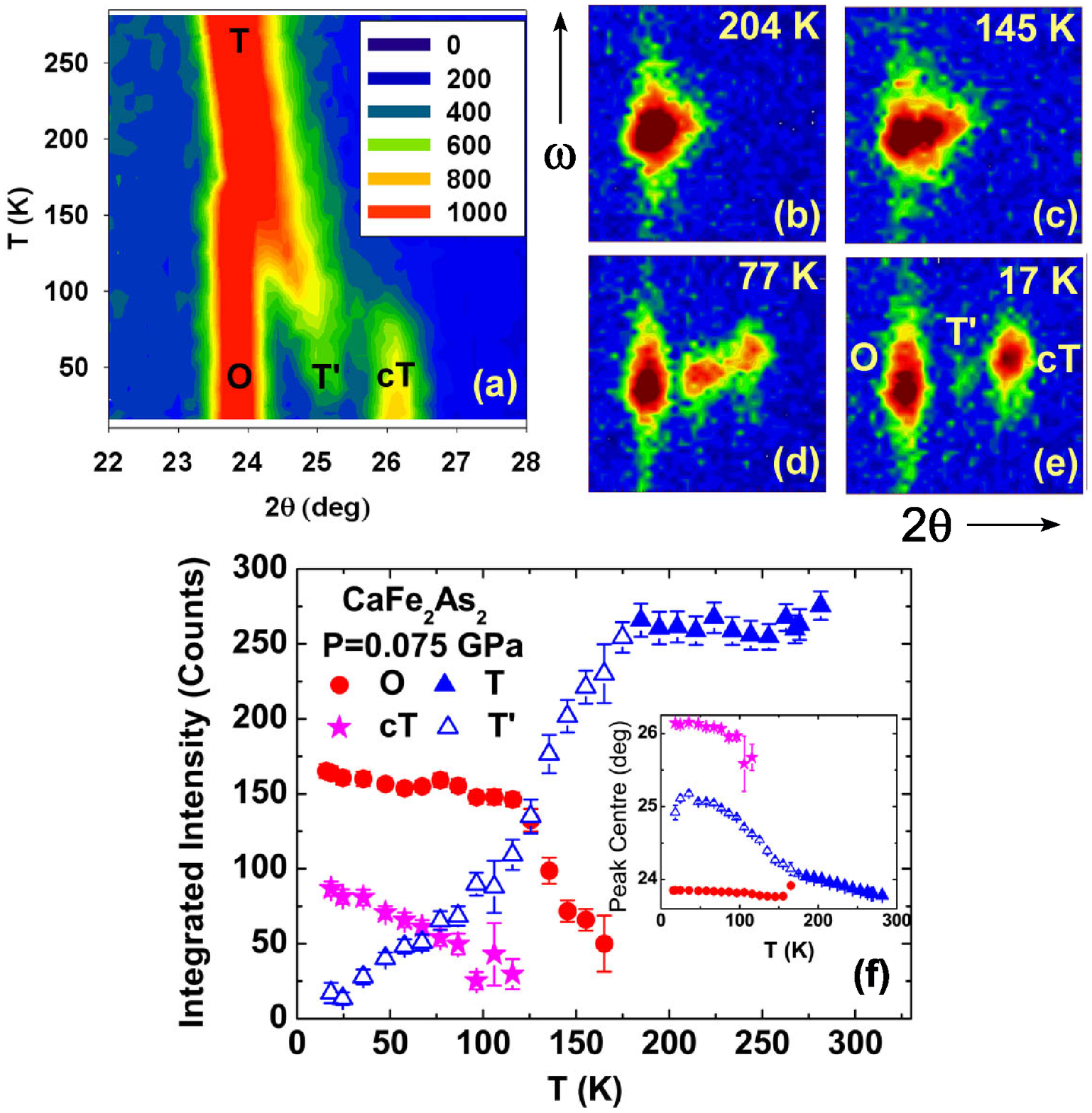}
\caption{(Color online) (a) Color map (counts per monitor are
color-coded in the inset) showing the temperature dependence of a
portion of the diffraction pattern taken on the E4 diffractometer
within the range of various structural (002) reflections of \cfa.
(b)-(e) $2\theta$-$\omega$ plots at selected temperatures showing
the angular distribution of peaks tracked in (a). Panel (f) shows
the temperature dependence of the integrated intensities and
positions of reflections shown in panels (a)-(e).} \label{Fig1}
\end{figure}

To investigate these issues we have performed single crystal neutron
diffraction measurements on \cfa~under uniaxial pressure. Since the
$c$-axis is subject to dramatic changes at the $T-cT$ transition,
the uniaxial pressure was applied along this direction in an attempt
to maximize the non-hydrostatic pressure component in a constrained
geometry. For pressures above 0.06~GPa we have observed diffraction
from a structure (which we initially labeled $T'$) that is
consistent with the stabilization of the high-temperature tetragonal
structure down to temperatures below the $SC$ transition. We also
find that with increasing applied pressures, the weight fraction of
the $T'$ and $cT$ phases increases at the expense of the $O$ phase
and the magnetically ordered fraction.  This identifies the $AF$
order with the $O$ structure, consistent with previous studies
\cite{Goldman:2009}. Finally, \emph{in-situ} measurements of the
in-plane ac resistivity (using the two-point contact method) clearly
reveal the onset of $SC$ below 10~K in our sample under uniaxial
pressure. Taken together, this observation suggests that SC is
hosted by the tetragonal phase which is stabilized under uniaxial
pressure.

Several high quality single crystals  of \cfa~with masses between
8-12 mg, and dimensions  of $\thickapprox$ 2-3mm x 3-4mm  x 0.2mm
were grown out of a Sn  flux as described previously
\cite{Ni,Kreyssig:2008a}. The crystals were  gently polished to
prepare flat and  parallel surfaces perpendicular to the $c$-axis.
Neutron Laue exposures  confirmed the good quality of samples after
the polishing  procedure. Crystals were subsequently clamped between
two ZrO$_2$ pistons that comprise a small uniaxial pressure cell
\cite{Kamarad:2008} capable of applying up to 1 kN of force on the
sample. The pressure  is  calculated from  the calibrated
displacement of the clamping screws and the  measured sample cross
section. We have investigated  five  different single crystals  at
several pressures  between ambient pressure and 0.3 GPa. It is
important to  note that the  force produced by Bellville  springs
acts along the $c$-axis of the sample, in  strong contrast to
 hydrostatic or quasi-hydrostatic experiments, and maximizes
the  possibility  to observe effects that were, in  the literature,
ascribed to non-hydrostatic conditions
\cite{Canfield:2009,Yu:2009,Torikachvili:2009a,Goldman:2009,Lee:2009}.
The pressure cell/sample system  represents  a confined  geometry
where thermal expansion and striction phenomena play an important
role. Therefore, all pressure values mentioned below refer to those
determined at room temperature. Perhaps most importantly, various
structural phases (e.g. the $T-cT$ transition) that result in
dimensional or volume changes cannot be regarded independently as
they  are mutually connected.

The neutron diffraction experiments were performed on the E4
double-axis diffractometer at the Helmholtz-Zentrum Berlin using a
neutron wavelength of $\lambda$=2.45~\AA\ and a standard cryostat.
Additional data sets and measurements of the magnetic diffraction
peaks were collected using the D10 diffractometer at the Institute
Laue-Langevin (ILL) with a wavelength of 2.36~\AA\ and a four-circle
closed cycle refrigerator capable of reaching temperatures down to
1.7 K. Both instruments make use of two-dimensional area detectors
that provide a diffraction image over a range of scattering angles
($2\theta$) as the sample is rocked over a specified angular range
($\omega$).  This considerably simplifies the task of mapping the
evolution of the scattering with temperature (see
Fig.~\ref{Fig1}(b-e)). Pyrolytic graphite filters were employed in
both sets of measurements to reduce the higher harmonic contents to
less that 10$^{-4}$ of the primary beam.

In Fig.~\ref{Fig1}(a) we show the temperature dependence of the
signal in the vicinity of the (002)$_T$ Bragg reflection measured on
the E4 instrument with decreasing temperature. The nominal pressure
applied along the $c$ axis was 0.075~GPa. The actual 2D scan
profiles recorded at 204~K, 145~K, 77~K and 17~K are shown in
Fig.~\ref{Fig1}(b-e). As temperature is lowered below $\thickapprox$
170~K, the discontinuous change in the position of the bulk of the
scattering to slightly lower scattering angle signals the $T-O$
transition \cite{Budko:2009}. However, besides the $cT$ scattering
signal at higher angle, there is a significant "tail" of scattering
between signals originating from the $O$ and $cT$ phases that
persists down to at least 17~K (labeled $T'$ in Fig.~\ref{Fig1}).
Below $\thickapprox$ 100~K, the intensity of the diffraction peak at
this intermediate scattering angle decreases as the intensity of
scattering from the $cT$ phase increases. Several features of the
$T'$ diffraction peak in Fig.~\ref{Fig1} are noteworthy: (a) It is
clearly distinguishable from both the $O$ and $cT$ phase peaks as
temperature decreases, marking it as a different as yet unknown phase that coexists with
the $O$ and $cT$ phases at low temperatures; (b) There is no
discernible discontinuity in the intensity or position of the $T'$
diffraction peak as it evolves from the higher temperature (002)$_T$
diffraction peak and; (c) the appearance and increase in the weight
fraction of the $cT$ phase and corresponding decrease in
fraction (peak intensity) of the $T'$ phase below $\thickapprox$
100~K is consistent with the temperature range of the $T-cT$
transition as measured by neutron diffraction under hydrostatic
pressure conditions \cite{Goldman:2009} and recent electrotransport
measurements under uniaxial stress \cite{Torikachvili:2009a}. We,
therefore, identify the $T'$ diffraction peak as (002)$_T$ arising
from some volume of the sample that has been stabilized in the high
temperature $T$ phase due to the uniaxial pressure conditions. It is
noteworthy that no trace of this stabilized $T'$ phase was observed
in neutron diffraction measurements under hydrostatic pressure
conditions \cite{Goldman:2009}.

Following the measurement on E4 we investigated freshly polished
\cfa~samples using the D10 instrument, which is equipped with a
4-circle stage, to extend these measurements to other crystal
orientations and characterize the magnetic scattering as well.
Measurements at several pressures confirmed the picture described
above including the emergence of the $T'$ and $cT$ phases as the
temperature was decreased below the $T-O$ transition.  We also
found, however, that this depends on the prior pressure history of
the sample. Below a starting pressure of $\thickapprox$ 0.06 GPa,
the $cT$ phase is absent at all temperatures and the $T'$ phase is
observed only over a narrow range of temperatures below $T_{TO}$.
Upon increasing the pressure, we observe that the relative
fractions of the stabilized tetragonal ($T'$) and $cT$ phases
increase at the expense of the O phase, as would be expected from
the $p-T$ phase diagram in reference \cite{Goldman:2009}. A finite
weight fraction of the $T'$ phase extends down to the lowest
temperatures measured for uniaxial pressures greater than 0.075~GPa.
At even higher applied pressure, the $cT$ phase appears at
progressively higher temperatures and its weight fraction increases
together with that of the $T'$ phase at the expense of the $O$
phase. For pressures in excess of $\thickapprox$ 0.27 GPa, the
relative fraction of T' decreases (see Fig. 3(c) inset). The lattice
constants of the $T'$ phase in the low temperature limit at 0.092
GPa were determined from the D10 data to be $a$ = 3.82 (8) {\AA}~and
$c$ = 11.44 (5) {\AA}. Since the data set was limited by strong
absorption of the cell and overlapping reflections from the O phase,
a full structural refinement of the T' phase could not be performed.

\begin{figure}
\includegraphics*[scale=0.31,angle=270]{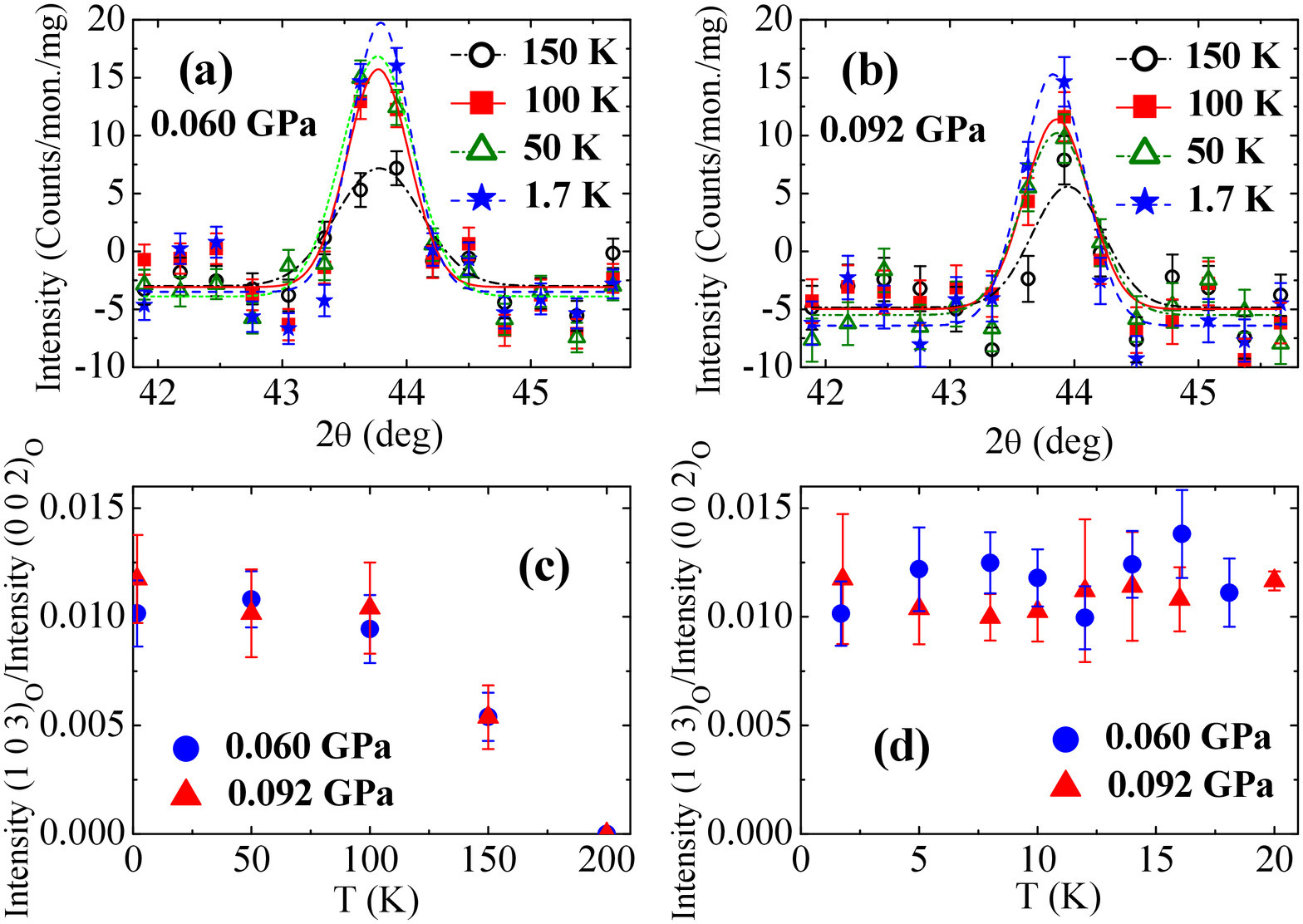}
\caption{(Color online) Representative examples of the magnetic
diffraction peak observed at the ($ 1\over {2}$ $1\over {2}$
${3}$)$_T$ (or (${1}$ 0 ${3}$)$_O$) position obtained from two
\cfa~samples using the D10 instrument under uniaxial pressures of
(a) 0.060~GPa and (b) 0.092~GPa applied along the $c$-axis
(normalized to the same monitor and crystal weight). Line through
the data represent fits using a single Gaussian. (c) The temperature
dependence of the integrated intensities of data shown in (a) and
(b) normalized to the intensity of the (002)$_O$ reflections at the
corresponding temperatures and pressures. (d) The normalized
integrated intensities of the magnetic reflections in the vicinity
of the expected superconducting transition (at $\thickapprox$ 10~K)
for $P$ = 0.060 and 0.092~GPa.} \label{Fig3}
\end{figure}

In Figs.~\ref{Fig3}(a) and (b) we show representative diffraction
profiles of the strongest magnetic reflection ($ 1\over {2}$ $1\over
{2}$ ${3}$)$_T$, (or (${1}$ 0 ${3}$)$_O$, in the $O$ unit cell
notation).  These data were taken on two \cfa~samples on the D10
instrument under uniaxial pressures of 0.060 GPa and 0.092 GPa
applied along the $c$ axis, respectively. For comparison, the
profiles were normalized to the same monitor and crystal weight.
Figs.~\ref{Fig3}(c) and (d) display the temperature and pressure
dependence of fits to the data shown in Figs.~\ref{Fig3}(a) and (b)
and reveal several important clues regarding the magnetism and
$SC$ in \cfa. First, we note that the normalized integrated
intensities taken at starting pressures of 0.060~GPa and 0.092~GPa
are essentially indistinguishable. Using the (002)$_O$ nuclear peak
for normalization, we estimate an ordered moment of 0.8(1) $\mu_B$
in good agreement with data in the literature. Second, within the
given sensitivity limit of about 0.2 $\mu_B$ we found no evidence of
magnetic ordering within the $T'$ phase at any of the temperatures
and pressures investigated. Together, this means that the magnetic
scattering intensity at each pressure is simply proportional to the
 fraction of the $O$ phase and there is no change in the
magnetic moment value in the $O$ phase with increasing pressure,
consistent with previous results from measurements under hydrostatic
pressure \cite{Goldman:2009}. We also point out that elastic\cite{Goldman:2009} and inelastic\cite{Pratt:2009b} neutron scattering measurements of under hydrostatic pressure have noted the absence of both a static ordered moment and low-energy spin-fluctuations in the $cT$ phase.
Finally, focusing our attention on
Fig.~\ref{Fig3}(d), we find no evidence of suppression of the
magnetic ordering at these pressures below the onset of SC as has
been observed, for example, in recent measurements on doped
$Ba(Fe_{1-x}Co_{x})_{2}As_{2}$ superconducting samples
\cite{Pratt:2009,Christianson:2009}. These data, then, are
consistent with the identification of the $AF$ ordering with the $O$
phase and the absence of $SC$ in the $O$ phase for these samples.
For the latter point, however, it is important to establish whether
these samples under applied uniaxial pressure are, indeed,
superconducting.

\begin{figure}
\includegraphics*[width=0.33\textwidth, height=0.45\textwidth]{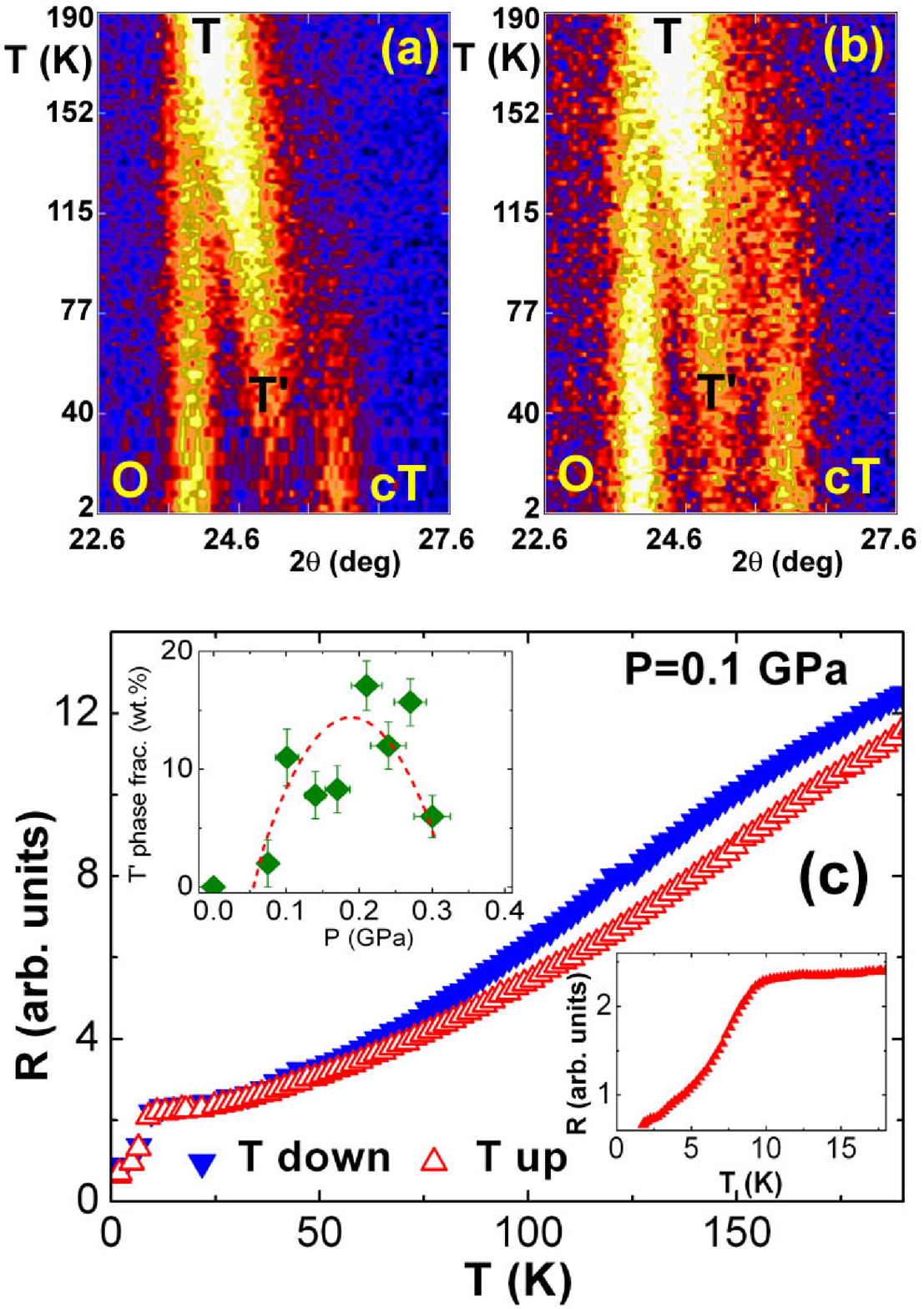}
\caption{(Color online) Color map showing the temperature dependence
of a portion of the diffraction pattern taken at E4 in that covers
various structural (002) reflections of \cfa~under uniaxial pressure
of 0.1 GPa along the $c$ axis measured with (a) decreasing and (b)
increasing temperature. The simultaneously measured electrical
resistance is shown in panel (c). The lower inset to (c) shows the
low-temperature detail of the electrical resistance data taken upon
heating. The upper inset shows the weight fraction of the pressure
stabilized phase $T'$ as a function of applied pressure. The dotted
line is a guide to the eye.} \label{Fig4}
\end{figure}

Figs.~\ref{Fig4}(a) and (b) display the temperature dependence of
the diffraction near the (002)$_T$ Bragg reflection measured on the
E4 instrument with (a) decreasing and (b) increasing temperature.
The nominal pressure applied along the $c$-axis in the present case
was 0.1~GPa. Simultaneously, we measured the temperature dependence
of the ac in-plane electrical resistivity using a two-point probe
and the results are shown in Figs.~\ref{Fig4}(c) with the
low-temperature detail magnified in the lower inset. From these
data, we see that the onset of $SC$ is clearly visible just below 10
K, although the resistivity does not reach zero even at 1.7 K. It is
well known, however, that the 2-point method always senses the
residual contact resistivity and the measured values are greater
than zero at all temperatures. From the relevant (002) reflection
intensities we estimate the weight fraction of the $T'$ phase at the
lowest temperature to be $\thickapprox$ 10 wt.\%. We note that the
resistivity curve does not exhibit any sharp anomaly near the $T-O$
transition and, overall, is reminiscent of the data taken under
uniaxial pressure ($\thickapprox$ 0.14-0.17~GPa) by Torikachvili et
al. \cite{Torikachvili:2009a}. By performing analogous diffraction
experiments at different applied pressures we have completed the
pressure dependence of the weight fraction of the $T'$ phase, which
is shown in the upper inset of Fig.~\ref{Fig4}(c). The onset of $SC$
\cite{Torikachvili:2009a} occurs coincident with the first appearance of the $T'$ phase.  Unfortunately, the maximum uniaxial pressure attainable in our measurements is below that required for the
offset of $SC$.

Summarizing our results on \cfa: For applied uniaxial pressures
above 0.060~GPa one induces the $cT$ phase which appears at
progressively higher temperatures, and the $T'$ phase that, for
applied pressures 0.075 GPa $<$ $P$ $<$ 0.3 GPa, is stabilized down
to the lowest temperature measured (1.7~K). We propose that the
critical factor for $SC$ in \cfa~in both uniaxial and frozen medium
pressure measurements is the stabilization of the tetragonal phase
at low temperatures. Our observations correspond very well with the
appearance of the superconducting dome as a function of the uniaxial
pressure as observed by Torikachvili et al.
\cite{Torikachvili:2009a}. The uniaxial pressure necessary to make
\cfa~superconducting is about an order of magnitude lower than
nominal "hydrostatic" pressure values produced by liquid medium
clamping cells, i.e. approximately of the same order of magnitude as
the non-hydrostatic component in the clamped cells. Although it is
conceivable that SC is hosted by some other, as yet undetected
additional phase, or through some strong modification of the $O$ or
$cT$ phase behavior that is not found under hydrostatic pressure
conditions, we view this as unlikely in light of the consistent
observation of SC in liquid media pressure cells independent of the
sample preparation methods. Finally, we note that the
superconducting "bubble" observed for \cfa~extends over only a
relatively narrow range of pressure (vanishingly small for the
hydrostatic pressure measurements). This is consistent with our
picture since with increasing pressure (at pressures where the $O$
phase is supressed) the fraction balance between the $T$ and $cT$
phases changes in favor of the $cT$ phase. At high enough pressures
the entire sample transforms to the non-superconducting $cT$ phase.

We gratefully acknowledge the HZB and ILL for the allocated beamtime
and their support for this work. The work at the Ames Laboratory was
supported by the U.S. DOE under Contract No. DEAC02- 07CH11358.


\end{document}